\documentclass{article}
\pdfoutput=1

\usepackage{arxiv}

\usepackage[utf8]{inputenc} 
\usepackage[T1]{fontenc}    
\usepackage{hyperref}       
\usepackage{url}            
\usepackage{booktabs}       
\usepackage{amsfonts}       
\usepackage{nicefrac}       
\usepackage{microtype}      
\usepackage{lipsum}		    

\RequirePackage[]{natbib}

\usepackage[american]{babel}
\usepackage{csquotes}
\usepackage{amsmath}
\usepackage{todonotes}
\usepackage{tikz}
\usepackage{float}
\usepackage{lineno}
\usepackage{multirow}
\usepackage{adjustbox}
\usepackage{amsthm}
\usepackage{amssymb}
\usepackage{graphicx}
\usepackage[figuresright]{rotating}
\usepackage{comment}

\usepackage{todonotes}

\title{Contextual aggregation and rapid updating of trial outcomes within a user-friendly open-source environment}

\author{
   František Bartoš$^1$, Eric-Jan Wagenmakers$^1$, Christiaan H. Vinkers$^{2,3}$, Kees P. J. Braun$^4$, and Willem M. Otte$^4$\\ \\
    \footnotesize{$^1$ Department of Psychological Methods, University of Amsterdam, The Netherlands}\\
    \footnotesize{$^2$ Department of Psychiatry, Amsterdam UMC, Vrije Universiteit Amsterdam, Amsterdam, The Netherlands}\\
    \footnotesize{$^3$ Amsterdam Public Health, Mental Health Program and Amsterdam Neuroscience}\\
        \footnotesize{Mood, Anxiety, Psychosis, Sleep \& Stress Program, Amsterdam, The Netherlands}\\
    \footnotesize{$^4$ Department of Pediatric Neurology, UMC Utrecht Brain Center,}\\
        \footnotesize{University Medical Center Utrecht, and Utrecht University, The Netherlands}
}

\renewcommand{\shorttitle}{Accessible aggregation and updating of evidence}

\begin{document}
\maketitle

\begin{abstract}
The delayed and incomplete availability of historical findings and the lack of integrative and user-friendly software hampers the reliable interpretation of new clinical data. We developed a free, open, and user-friendly clinical trial aggregation program combining a large and representative sample of existing trial data with the latest classical and Bayesian meta-analytical models, including clear output visualizations. Our software is of particular interest for (post-graduate) educational programs (e.g., medicine, epidemiology) and global health initiatives. We demonstrate the database, interface, and plot functionality with a recent randomized controlled trial on effective epileptic seizure reduction in children treated for a parasitic brain infection. The single trial data is placed into context and we show how to interpret new results against existing knowledge instantaneously. Our program is of particular interest to those working on the contextualizing of medical findings. It may facilitate the advancement of global clinical progress as efficiently and openly as possible and simulate further bridging clinical data with the latest biostatistical models.
\end{abstract}


\section{Introduction}
Proper interpretation of novel clinical trial results requires contextual embedding. Single trials emerge from and build upon existing knowledge. Reinforcement with previous information solidifies the newly reported outcomes into valid evidence. In evidence-based medicine, this contextual approach of solidifying a single trial’s findings within its broader context is traditionally operationalized as meta-analytic aggregation and summarization. 

In an ideal world, this assessing and placing into the perspective of a particular disease state, treatment, or outcome is done instantaneously. In practice, it requires compiling an overview paper with access to the historical trace of previous trials investigating the same outcome as the novel trial. It also requires reliable analysis within a methodically up-to-date modeling framework.

These two elements -- historical data and modeling -- are ideally incorporated and accessible into a single program if we strive for a rapid translation from research to clinic. Less obvious but of equal importance would be the accessibility of this program to everyone with an interest in working according to the evidence-based medicine principle, namely to incorporate all of the latest information in clinical decision-making and treatment strategies. Despite active endeavors to provide all physicians and scholars with open and free access to the latest clinical trial information, boundaries such as paywalls still hinder clinicians and healthcare policymakers working outside university-licensed information networks.

Therefore, we need an open, regularly updated, large, standardized dataset of previously published clinical trial outcomes to offer a clinically helpful historical perspective covering the most critical and active medical research field. We need to connect this dataset with an open, user-friendly software environment to integrate it with reliable and methodically up-to-date modeling. An environment that is sufficiently flexible to keep incorporating the latest tools and insights from the biostatistical community provides robust and sound aggregations and informative visualizations.

We aimed to collect a representative dataset and combine it with newly developed meta-analytical models, accessible through a user-friendly and intuitive interface; an interface allowing users to select the appropriate historical trial results as well as results from a novel clinical trial, not in the database. We also aimed to provide the latest validated classical and Bayesian meta-analytical models and straightforward summarization plots.

To achieve this goal, we extracted clinical trial data from high-quality clinical overview papers, covering an eclectic amount of the approximately 1.5 million published trials to date. We implemented a frequentist and Bayesian meta-analytic pipeline for various types of outcome data (e.g., continuous, dichotomous) and demonstrated the program based on a recent emerging evidence case in the field of global health/neurology.

\section{Methods}

\subsection{Database construction}

We based our clinical trials database on data extracted from the Cochrane Database of Systematic Reviews (CDSR). These reviews cover all medical fields, have high-quality standards and methodological rigor with elaborate search protocols, and rigorously identify and summarize comparable trials \citep{jorgensen2006cochrane}. Moreover, these reviews perform meta-analyses on individual clinical trials to generate an estimated effect size of interventions. 

We identified all systematic reviews in the CDSR through PubMed with the NCBI’s EUtils API (query: "Cochrane Database Syst Rev"[journal] AND ("2000/01/01"[PDAT]: "2021/01/31"[PDAT]). We downloaded the XML meta‐analysis table file (rm5‐format) associated with the review’s latest version. We extracted limited meta-information (keywords, study title) for each clinical trial in a meta-analysis table, including the first author’s name, year, and raw outcome numbers. We discarded all additional information, such as Cochrane’s model estimates and subgroup analyses. All data were converted to a single file in plain text format.

\subsection{Software environment}

We implemented our framework as a module in JASP \citep{JASP16, ly2021bayesian}. JASP is a free and open-source program for statistical analysis designed to be intuitive for non-statistical academic scholars. In the implementation, we offer meta-analysis functionality in both their classical and Bayesian form, so frequentist inference and Bayesian inference using the same meta-analytical models. Frequentist inference uses p-values and confidence intervals to control error rates in the limit of an infinite number of perfect replications. Bayesian inference uses credible intervals and Bayes factors to determine credible parameter values and quantify model evidence given the available data and prior knowledge. In addition, we offer more recent meta-analytical approaches, including fixed and random effects averaging (Bayesian only) and incorporation of prior information (Bayesian only). With our dedicated visualization of diagnostic data and aggregated estimates, we facilitate the rapid comparison between different approaches.

\subsection{Efficient data storage and search functionality}

To handle 1.5 million trial outcomes efficiently, the meta-data and database are stored in separate rds files (for both the continuous and dichotomous outcomes). We use the meta-data to dynamically generate the responsive graphical user interface. The interface allows users to generate a list of reviews/meta-analyses based on either topic, keywords, or a text-based title search to extract the relevant historical insights. Users select the desired meta-analysis (and corresponding subgroups if available) from the generated list, choose which group from each meta-analysis is treated as the comparison group, and decide whether they want to analyze the meta-analyses individually or pooled together. In addition, users can manually extend the included database through the graphical user interface to include results from new studies.

In case of dichotomous outcomes, the users can further decide whether they are interested in analyzing the data as log odds ratios, log Peto’s odds ratios, log risk ratios, or risk differences, with all effect size calculations handled internally by JASP. 

\subsection{Frequentist fixed effect, random effects, and model-averaging features}

After users specify the desired meta-analyses and selection, the meta-data is paired with the database which is forwarded to the JASP meta-analytic functionality. Our framework relies on JASP’s classical and Bayesian meta-analysis modules (e.g., \citealt{berkhout2021tutorial}) which are mainly built upon the \texttt{metafor} \citep{metafor} and \texttt{metaBMA} \citep{metaBMA} \texttt{R} \citep{R} packages. Both modules offer state-of-the-art evidence synthesis features.

For example, the classical meta-analysis offers all the usual features such as various meta-analytic estimators, visualizations, and model diagnostics tools. The Bayesian meta-analysis implements both the estimation and Bayes factor hypothesis testing (e.g., \citealp{jeffreys1935some, kass1995bayes}) with fixed and random effects models and also the novel Bayesian model-averaged meta-analysis (BMA). BMA combines the aforementioned fixed and random models according to their predictive performance and weighs their inference appropriately (see \citealp{gronau2021primer, bartos2021bayesian} for more detail). In addition, users can specify informed prior distributions and generate a wide range of model visualizations.

Our framework also allows users to export the selected studies as universally readable csv files. Consequently, the selected studies can be re-analyzed with an even richer set of features provided in JASP, e.g., both the classical and Bayesian publication bias adjustment techniques \citep{bartos2020adjusting}, or any other statistical software.

\section{Case Example: Neurocysticercosis}

We demonstrate the program based on a recent trial published on a treatment for neurocysticercosis \citet{singh2022efficacy}. Human cysticercosis, an infection caused by the infestation of the larval stage of the zoonotic pork tapeworm Taenia solium, is a severe but neglected zoonotic disease and a serious public health problem in low- and middle-income countries \citep{coyle2012neurocysticercosis}. People with an intestinal tapeworm disseminate infectious eggs in their feces, contaminating vegetables and food in areas with poor sanitary conditions. If ingested, these eggs can cause cysticercosis in both humans and pigs. The infection-induced calcification of the cysts is a major cause of acquired epilepsy and the single most important reason the incidence of epilepsy in the developing world is significantly higher compared to high-income countries. The median incidence of epilepsy in high-income countries is an estimated 45.0/100,000/year (interquartile range [IQR] 30.3–66.7) compared to 81.7 (IQR 28.0–239.5) for low- and middle-income countries \citep{ngugi2011incidence}. The most common clinical manifestation of cysticercosis is focal seizures.

Treatment with anthelmintics probably hastens the radiological resolution of cysts (risk ratio 95\% confidence interval (CI): $1.07$--$1.39$) and may slightly reduce the risk of developing epilepsy (i.e., recurrent seizures, CI: $0.78$--$1.14$). However, the beneficial effect is contested if more than one cyst is present \citep{monk2021anthelmintics}. Nearly forty percent of infected brains show residual cysts at six months follow-up after treatment, resulting in higher long-term epilepsy risks. Anthelmintics could also lead to adverse drug reactions \citep{bagheri2004adverse}, so more evidence in favor of treatment is welcome. This evidence includes answering the question if combining two anthelmintics is helpful, as it may potentially result in more side effects and adverse drug reactions but also lead to higher levels of radiological resolution and lower seizure rates.

Recently \citet{singh2022efficacy} conducted a randomized double-blinded study with three groups of children treated for neurocysticercosis: one combining the anthelmintics albendazole and praziquantel therapy, one with albendazole monotherapy and one with a placebo treatment.

Their group sample sizes are small, with twenty-one children on a combination of albendazole and praziquantel, nineteen children on  albendazole-only treatment, and twenty receiving a placebo. All treatments were given for one month. At three- and six-month follow-ups, the brain lesion’s resolution, and seizure-recurrence were characterized. An important question is whether the recurrence characteristics hold if this new trial is placed against the background of previous data on the same treatment. 

The clinical outcome was measured in terms of seizure recurrence. Two children encountered seizure recurrence in the first month starting treatment: one in the placebo group and one in the albendazole-only group. No recurrent seizures were reported in the albendazole and praziquantel combination group. The authors concluded that ``The combination therapy did not result in increased seizure recurrence or adverse drug reaction compared with albendazole monotherapy'' (p. 371)  but also acknowledged that the single-center study was limited by ``patient enrollment of modest magnitude'' (p. 371).

The seizure recurrence in both the albendazole alone and placebo groups raises the question of the efficacy of albendazole monotherapy in children. In routine care settings praziquantel is not as available as albenazole and may limit combination therapy. How should we weigh the two cases of recurrent seizures against the backdrop of previous trials characterizing seizure recurrence in children during albendazole alone versus placebo treatment?

We demonstrate how our program can address this question in a series of steps. First, we show the database selection mechanisms, perform a simple re-analysis of the seizure-recurrence, and explain the main output and options. Second, we repeat the analysis but select only the children subgroup of participants and extend the results with new findings. Third, we show the fixed and random effects model-averaging on the seizure-recurrence using the Bayesian model-averaged meta-analysis.

\subsection{Re-analysis of seizure-recurrence in children}
First, we show how to load and re-analyze the \citep{monk2021anthelmintics} meta-analysis of albendazole on neurocysticercosis. We open JASP and use the additional module selection menu denoted by a large ``+'' on the top right corner of the program. We load the ``Cochrane Meta-Analyses'' module by selecting the corresponding checkbox which adds the module icon on the main ribbon. We click the icon and select the ``Classical: Dichotomous Outcomes'' analysis.

\begin{figure}[h]
    \centering
    \includegraphics[width=\textwidth]{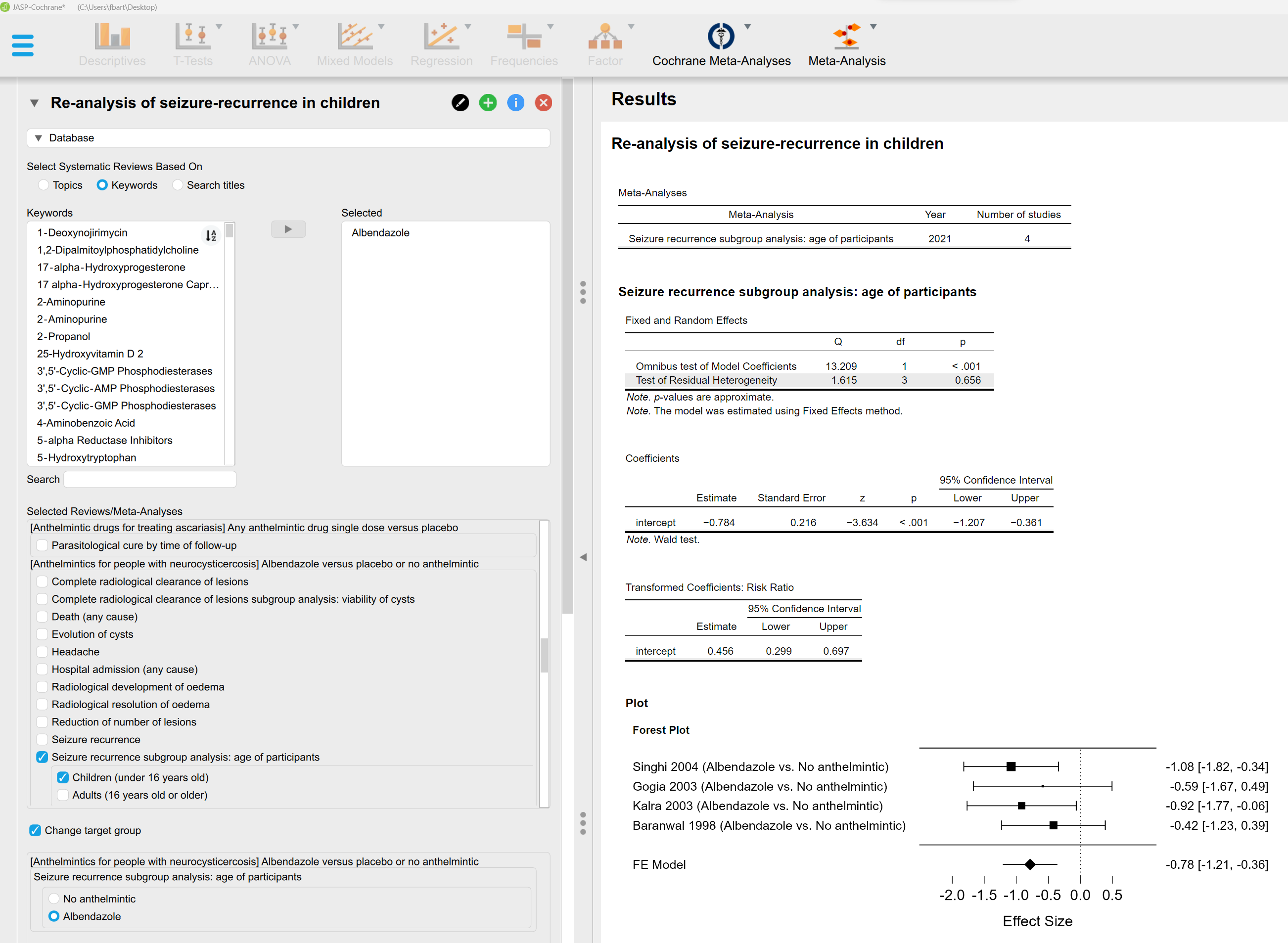}
    \caption{Screenshot from JASP Cochrane Meta-Analysis module. Left panel: the graphical user interface for interacting with the database; right panel: the resulting meta-analytic output based on the \citet{monk2021anthelmintics} data.}
    \label{fig:JASP-1}
\end{figure}

\subsubsection{Selecting the dataset}
Figure~\ref{fig:JASP-1} shows JASP's graphical user interface of the analysis for the example data. The left panel contains the ``Database'' section of the user interface (shared across the Cochrane Meta-Analysis module). The database section is generated from the metadata and used for selecting the meta-analyses. Here we used the ``Keywords'' option to filter the provided reviews based on keywords.\footnote{The ``Topics'' option would filter the reviews based on Cochrane database assigned keywords, and the ``Search titles'' option would filter the reviews on text search of titles.} We find the ``Albendazole'' entry and move it to the ``Selected'' box. Consequently, the ``Selected Reviews/Meta-Analyses'' box is populated with a set of meta-analyses listed within the corresponding reviews. We scroll through the list until we find ``[Anthelmintics for people with neurocysticercosis] albendazole versus placebo or no anthelmintic'' review. We select the ``Seizure recurrence subgroup analysis: age of participants''. After ticking the checkbox, two new checkboxes appear underneath the selected meta-analysis. We deselect the ``Adults (16 years old or older)'' option and proceed only with the ``Children (under 16 years old)'' data. We change the target group to ``Albendazole'' (to obtain results in the same direction as reported in CDSR).

\subsubsection{Analysis options and output}
Immediately after selecting the data set, the output panel of JASP (i.e., the right panel from Figure~\ref{fig:JASP-1}) presents a summary of the selected dataset: the name of the meta-analysis, the year of publication, and the number of studies. JASP also automatically generates the default meta-analytic results for each selected meta-analysis (unless we wish to analyze the pooled data; see below). To obtain results similar to the CDSR report, we changed the effect size scale to log risk ratios (logRR) and selected the fixed effect method.\footnote{The CDSR report uses Mantel-Haenszel fixed method which leads to highly similar results.} The ``Fixed and Random Effects'' table summarizes the tests for heterogeneity, showing that the four included studies do not prompt the rejection of the null hypothesis of no heterogeneity -- a point we address with BMA in the last section. 

The coefficients table summarizes the meta-analytic estimate on the logRR scale, $\text{logRR} = -0.784$, CI: $-1.207$--$-0.361$, $z = -3.634$, $p < 0.001$, and the transformed coefficients table automatically transforms the estimate to the risk ratio scale, $\text{RR} = 0.456$, CI: $0.299$--$0.697$. These results show a lower risk of seizure-recurrence in the albendazole-treated group. To generate the forest plot summarizing the individual studies and the meta-analytic estimate, we open the `Statistics'' section of the interface and select the appropriate checkbox (see the left panel of Figure~\ref{fig:JASP-2}). Furthermore, we see a range of additional options, such as fit measures, funnel plots, or funnel plot asymmetry tests, all accessible with a single mouse click.

\subsection{Incorporating new findings}
\begin{figure}[h]
    \centering
    \includegraphics[width=\textwidth]{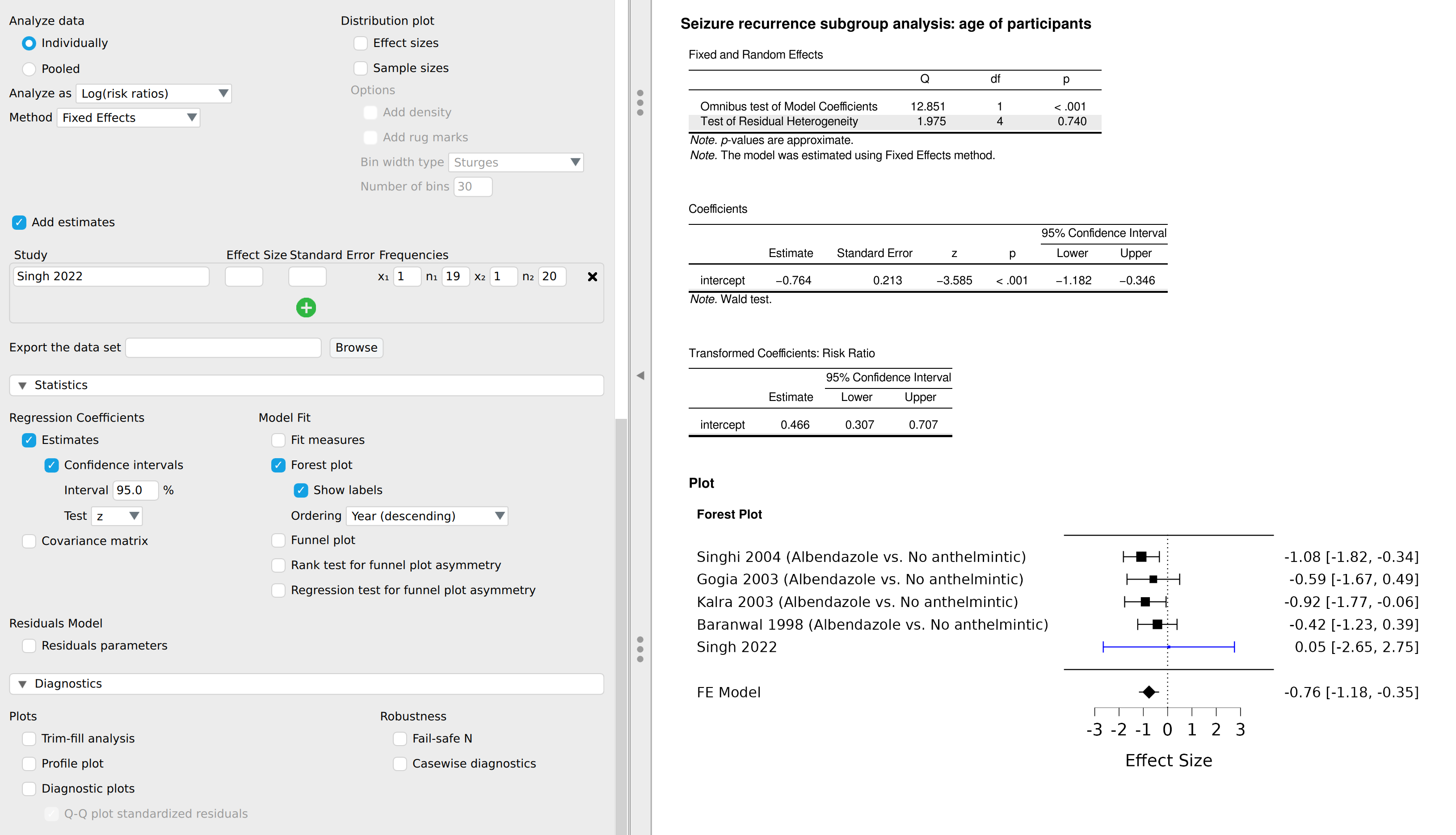}
    \caption{Screenshot from the JASP Cochrane Meta-Analysis module highlighting, in the left panel, the graphical user interface for incorporating new findings and requesting additional Statistics and Diagnostics options. The right panel displays the resulting output from extending the \citet{monk2021anthelmintics} meta-analysis with the results from \citet{singh2022efficacy}.}
    \label{fig:JASP-2}
\end{figure}

Now we extend the results with the data from \citet{singh2022efficacy}. The left panel of Figure~\ref{fig:JASP-2} shows the user interface for adding new findings, which appears after selecting the ``Add estimates'' checkbox. We can enter the study name, ``Singh 2022'', and either supply an already computed effect size estimate and its standard error, or input the observed frequencies in each group. We proceed with the latter option and complete the $x_1 = 1$, $x_2 = 1$, $n_1 = 19$, and $n_2 = 20$ fields where $x$ corresponds to the number of events and $n$ to the number of observations for the first and the second group, respectively (note that we set the target group, i.e., first group, to albendazole in the previous step).

Having entered the data via the interface, the analysis then refreshes automatically and generates results that include the new findings (i.e., the blue effect size estimate labeled ``Singh 2022'' has been added to the forest plot). Note that, first, the logRR estimate from \citealt{singh2022efficacy} is accompanied by a confidence interval that is relativelty large (i.e., $\text{logRR} = -0.05$, CI: $-2.65$--$2.75$), indicating that the estimate is relatively uncertain. Second, note that the updated meta-analytic estimate, $\text{RR} = 0.466$, CI: $0.307$--$0.707$, $z = -3.585$, $p < 0.001$, remains essentially unchanged and still shows a lowered risk of seizure-recurrence in the albendazole group in comparison to placebo in children. It appears that in the face of already existing knowledge, the small sample results from the \citet{singh2022efficacy} should not change our conclusions about the efficacy of albendazole in reducing seizure-recurrence in children. 

\subsection{Bayesian model-averaged meta-analysis of seizure-recurrence}
Finally, we analyze the seizure-recurrence data set extended with the \citet{singh2022efficacy} data from a Bayesian angle. There are multiple advantages (but also added difficulties) that come with a Bayesian analysis. Here, we illustrate how to go beyond statistical significance and directly quantify evidence in favor, or against, the null hypothesis, and effortlessly perform a Bayesian model-average meta-analysis that combines results from both the fixed and random effects models.

In contrast to the previous section, we select the ``Bayesian: Dichotomous Outcomes'' analysis from the Cochrane Meta-Analysis module. Then, as in the previous section, we use the `Database` section to select the ``Seizure recurrence'' meta-analysis and again change the target group, effect size scale, and add the \citet{singh2022efficacy} results.

\begin{figure}[h]
    \centering
    \includegraphics[width=\textwidth]{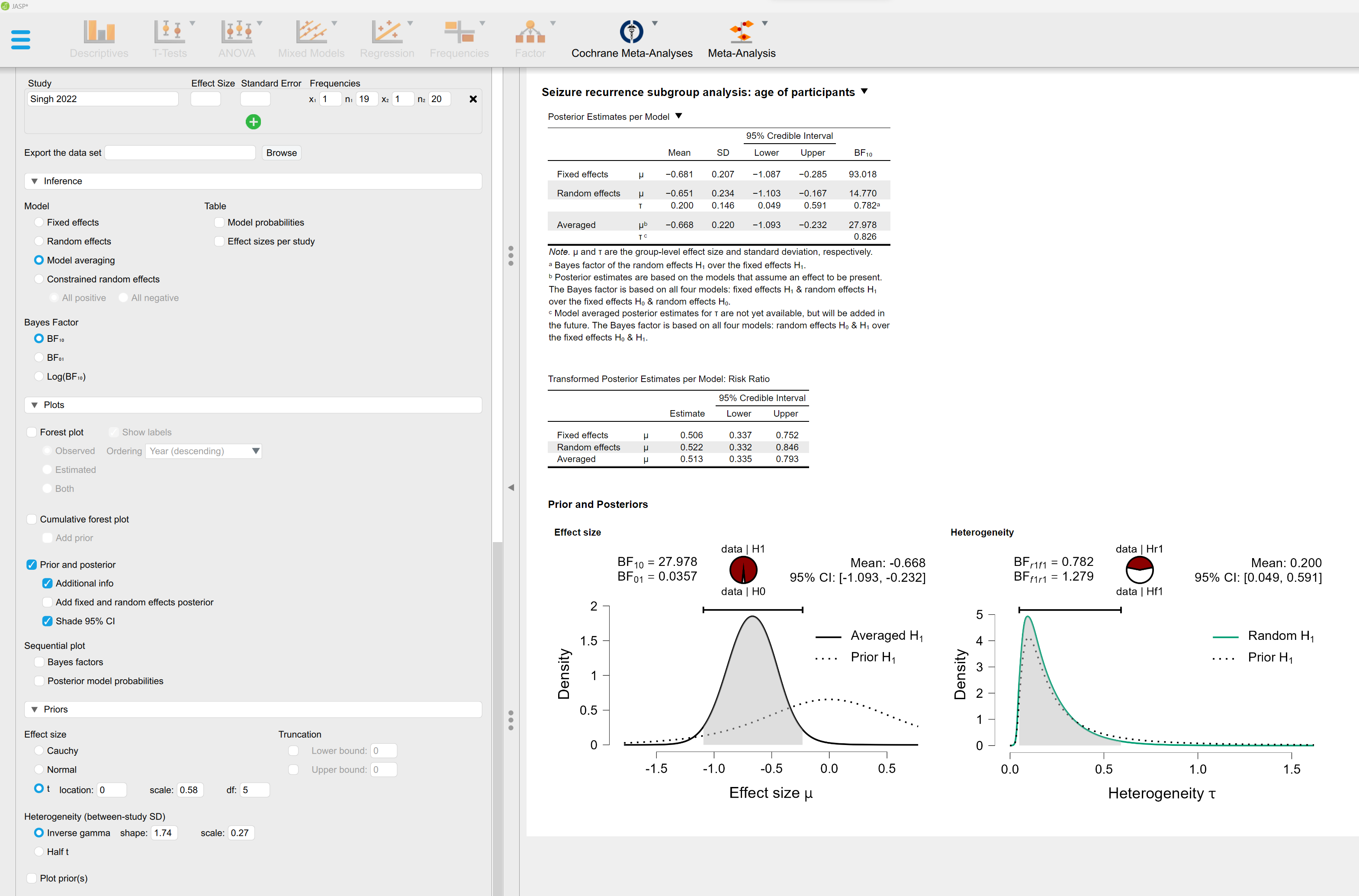}
    \caption{Screenshot from JASP Cochrane Meta-Analysis module highlighting, in the left panel, the graphical user interface for specifying Bayesian meta-analyses, model visualization, and prior distribution specification. The right panel displays the resulting output from extending the \citet{monk2021anthelmintics} meta-analysis with the results from \citet{singh2022efficacy}.}
    \label{fig:JASP-3}
\end{figure}

\subsubsection{Specifying prior distributions}

The primary challenge in performing a Bayesian analysis lies in specifying an alternative hypothesis via a prior distribution on the model parameters. The additional work is, however, rewarded with the ability to evaluate the evidence that the data provide regarding the comparison between the null hypothesis versus the alternative hypothesis. Evaluation of evidence in favor of either of the hypotheses qualitatively differs from the classical settings where we compute the probability of encountering data at least as extreme as those observed if the null hypothesis were true (i.e., the $p$-value).

The important point is that the specified prior parameter distributions quantify the predictions from the alternative hypothesis we are interested in testing. Here, we specify an informed prior distribution for both the effect size ($\mu$) and heterogeneity ($\tau$) parameters based on the Cochrane database itself \citep{bartos2021bayesian, bartos2023empirical}. Specifically, we use the prior distributions for the ``Epilepsy'' subfield and logRR from \citet{bartos2023empirical} and specify it in the ``Prior'' section; $\mu \sim \text{Student-t}(\text{location} = 0, \text{scale} = 0.58, \text{df} = 5)$ and  $\tau \sim \text{Inv-Gamma}_(\text{shape} = 1.74, \text{scale} = 0.27)$. These prior distributions can represent our expectations of small effects, i.e., we assign 71\% probability to the meta-analytic effect size to be within the $0.5$ and $2$ risk ratio interval, and meta-analytic heterogeneity to have mean of logOR = $0.36$. See \citet{bartos2021bayesian} for prior distributions for continuous outcomes, \citet{bartos2023empirical} for prior distributions for binary and time to event outcomes, and \citet{mikkola2021prior} for an overview of prior elicitation procedures.

\subsubsection{Analysis options and output}
After specifying the informed prior distributions, JASP automatically recomputes the analysis results. Because we retained the default ``Model averaging'' Model option in the ``Inference'' section, the right panel of Figure~\ref{fig:JASP-3} shows the ``Posterior Estimates per Model'' table. The first four columns contain the parameter estimates under the fixed effect model (effect size estimate only), the random effects model (effect size and heterogeneity estimate), and the model-average ensemble (effect size estimate only). The fixed and the random effects models produce comparable mean effect size estimates of albendazole on reducing the risk of seizure recurrence in children (logRR = $-0.681$ vs. logRR = $-0.651$), however, the fixed effect model results in approximately six times more evidence in favor of the presence of the effect than the corresponding random effects model ($\text{BF}_{10} = 93.02$ vs. $\text{BF}_{10} = 14.77$). This means it becomes important to address the question whether our inferences ought to be based on the fixed effect model or on the random effects model. One possible way is to have this decision be guided by the evidence for the absence vs. absence of heterogeneity. However, under the assumption of the presence of the effect, there is only weak evidence in favor of the absence of heterogeneity, $\text{BF}_{\text{rf}} = 0.782$, and this result therefore does not offer strong guidance for making an all-or-none decision.

An alternative solution is provided by Bayesian model-averaging. This procedure allows us to combine the parameter estimates and the evidence from both models according to their predictive performance (i.e., how well the models account for the data). In our case, the model-averaged effect size estimate (assuming the presence of the effect), is a data-determined compromise between the estimates from the fixed and the random effects model (i.e., $\text{logRR} = -0.668$). Furthermore, we obtain a (model-averaged) inclusion Bayes factor for the presence vs. absence of the effect, $\text{BF}_{01} = 27.98$, indicating strong evidence for the presence of the effect. The ``Transformed Posterior Estimates per Model'' table then summarizes the effect size estimates transformed to the risk ratio scale: the model-averaged estimate of the benefitial effect of albendazole on reducing the risk of seizure-recurrence in children, $\text{RR} = 0.513$, CI: $0.335$--$0.793$.

We conclude the example by visualizing the prior and posterior distributions of the effect size parameter by selecting the corresponding checkbox in the ``Plots'' section. The resulting plot (bottom right part of Figure~\ref{fig:JASP-3}) shows the prior distribution (common to all models) as a dashed line, and the model-average posterior distribution in black (with the shaded area corresponding to a central 95\% posterior credible interval). The data are informative for the effect size parameter: compared to the informed prior distribution, the model-averaged posterior distribution has shifted and is more narrow. However, the data are not informative for between-study heterogeneity: the informed prior distribution is almost identical in shape to the posterior distribution. This lack of information in the data highlights the importance of Bayesian model-averaging across the fixed and random effects models, as each provides a different degree of evidence regarding the presence of the treatment effect, but both models account for the data about equally well.

%
%
%



\section{Discussion}

The volume of clinical output is growing rapidly in the current evidence-based medicine. The speed of this growth constitutes a challenge for the creation of practical overviews and evidence-based synthesis. It also leads to difficulties in systematically updating knowledge. To help practicioners navigate the deluge of data we implemented an open framework that can accelerate the evaluation of new individual clinical trial results. By presenting powerful statistical models in a user-friendly, responsive interface we aim to facilitate cross-talk among active physicians who wish to use the latest biostatistical field insights. The proposed software solution may also be used to assist educational programs in the teaching of effective evidence collection and aggregation.

First, we focused on seizure-recurrence in children with neurocysticercosis and found a statistically significant effect of albendazole in reducing the risk of seizure-reccurence over placebo. Second, we highlighted how the current evidence can be extended by including data from a new trial. Finally, we performed an alternative analysis from a Bayesian perspective. Bayesian meta-analysis of the seizure-recurrence data revealed the difficulty of the classical all-or-none dichotomy (and user's choice) between a fixed effect or a random effects meta-analytic model: while the effect size estimates under both models were of comparable magnitude, the fixed effect model provided considerable more support for the alternative hypothesis. In order to resolve the all-or-none selection dilemma in a principled manner, we applied Bayesian model-averaging -- a procedure that provides a data-driven compromise between the fixed and random effect models to yield a single effect size estimate and a single Bayes factor.

\subsection{Strengths and limitations}
Our approach is unique as it combines a large set of historical data with the latest classical and Bayesian meta-analytical insights operationalized within an easy-to-use computer program. Other software is available to perform a meta-analysis. For example, Cochrane’s ReviewManager (latest version 5.4.1, Sep 2020) is a free standalone desktop tool reproducing frequentist fixed effect and random effects meta-analysis associated with a systematic review. However, only 46.8\% (7,483 out of 15,993, Oct 7th 2022)  of the systematic reviews provide free access to the full text and meta-analysis tables, limiting the approach's generalizability.

For a time-constrained physician or policymaker, it is generally not feasible to use R to apply the various meta-analytic tools outlined in this manuscript. However, our software tool makes effective use of the latest R packages wrapped inside a user-friendly environment, bringing state-of-the-art methodology directly to physicians and policymakers (without the need of writting, debugging, or documenting code).

Our program has limitations. Most importantly, we only incorporated all trial outcomes in systematic reviews published in the Cochrane Database of Systematic Reviews. Much more systematic reviews with meta-analyses are available, capturing clinical evidence in partly different trials and somewhat different contexts. The non-standardized presentation of these data makes it challenging to work with. It would require semi-automatic table extraction and reference matching based on PDFs or web versions of published meta-analyses. The varying quality of systematic reviews would also raise the issue of which meta-analyses to exclude from the database. Another limitation is the need to update the current dataset every time a major release of Cochrane systematic reviews is available. Automatic updating is not yet available but is feasible given the standardized way the meta-analytic data tables are stored and presented (i.e., in XML format).

\subsection{Final thoughts}
We believe that our program is helpful for general physicians and medical specialists. These medical professionals have limited time to study all systematic reviews in their field of expertise. It also benefits a much broader audience, including journalists and patients. It may also help empower patients, as anyone can use the program to bring isolated results into a broader perspective. Developing this tool for contextual evaluation of medical findings is a timely endeavor, given the increasing emphasis on achieving reliable information in a setting where many individual studies are still highly underpowered and potentially biased towards positive findings. Our interface may increase awareness of the importance of replication studies and funding schemes. Society is not only in need of novel insights. Equally important is the establishment of truth -– a truth that furthers clinical progress. Access to our evidence-calibrating model will enhance research in the rapidly growing and promising domain of statistical estimations and predictions. Our combinatory and open approach may help and stimulate similar projects and developments. 

We hope our program will play a role in new trial preparation. For example, the rapid access to historical trial data may allow for prior effect-size calculations and, subsequently, for determining how many subjects should be included in new trials. Ethical committees may also use the tool to appraise the proposed statistics in forthcoming trials. The increasing massive bulk of medical information makes it difficult to obtain the latest overview. Our program’s is potentially helpful in effective unlocking of existing knowledge for those making decisions on initiating new clinical trials.

\section{Acknowledgements}

This work was supported by The Netherlands Organisation for Scientific Research (NWO) through a Vici grant (to EJW; 016. Vici.170.083) and an NWA Idea Generator grant (to WMO; NWA.1228.191.045).

\section{Financial Disclosure}
None.

\section{Data Availability Statement}
The program’s software and database are publicly available at: \url{https://github.com/jasp-stats/jaspCochrane/}. JASP is available for download at \url{https://jasp-stats.org/} (current version 0.17.1). The code is accessible at: \url{https://github.com/jasp-stats/}.

\section{Potential Conflict of Interest}
František Bartoš and Eric-Jan Wagenmakers declare their involvement in the open-source software package JASP (https://jasp-stats.org), a non-commercial, publicly-funded effort to make Bayesian statistics accessible to a broader group of researchers and students. Christiaan H. Vinkers and Willem M. Otte are co-founders of DeepDoc (https://deepdoc.io), an AI platform providing weekly clinical trial notifications.

\bibliographystyle{biometrika}
\bibliography{manuscript.bib}

\end{document}